\documentclass{pasj00}
\def\inpress{in press}
\def\astroph#1{ (astro-ph/#1)}

\DeclareAbbreviation\AAHam{Astron. Abh. Hamburg. Sternw.}
\DeclareAbbreviation\AARv{Astron. Astrophys. Rev.}
\DeclareAbbreviation\an{Astron. Nachr.}
\DeclareAbbreviation\AcA{Acta Astron.}
\DeclareAbbreviation\Afz{Astrofizika}
\DeclareAbbreviation\AnTok{Tokyo Astron. Obs. Annals, Sec. Ser.}
\DeclareAbbreviation\Ap{Astrophysics}
\DeclareAbbreviation\ARep{Astron. Rep.}
\DeclareAbbreviation\ATel{Astronomer's Telegram}
\DeclareAbbreviation\ATsir{Astron. Tsirk.}
\DeclareAbbreviation\AcApS{Acta Astrophys. Sinica}
\DeclareAbbreviation\AstL{Astron. Letters}
\DeclareAbbreviation\BaltA{Baltic Astron.}
\DeclareAbbreviation\BASI{Bull. Astron. Soc. India}
\DeclareAbbreviation\BeSN{Be Star Newsletter}
\DeclareAbbreviation\GCN{GCN}
\DeclareAbbreviation\ibvs{Inf. Bull. Variable Stars}
\DeclareAbbreviation\JAD{J. Astron. Data}
\DeclareAbbreviation\JAVSO{J. American Assoc. Variable Star Obs.}
\DeclareAbbreviation\JBAA{J. British Astron. Assoc.}
\DeclareAbbreviation\LowOB{Lowell Obs. Bull.}
\DeclareAbbreviation\MitVS{Mitteil. Ver\"{a}nderl. Sterne}
\DeclareAbbreviation\MmSAI{Mem. Soc. Astron. Ita.}
\DeclareAbbreviation\Msngr{Messenger}
\DeclareAbbreviation\NewA{New Astron.}
\DeclareAbbreviation\NewAR{New Astron. Rev.}
\DeclareAbbreviation\OAP{Odessa Astron. Publ.}
\DeclareAbbreviation\Obs{Observatory}
\DeclareAbbreviation\PASA{Publ. Astron. Soc. Australia}
\DeclareAbbreviation\PAZh{Pis'ma AZh}
\DeclareAbbreviation\PhR{Phys. Rep.}
\DeclareAbbreviation\PVSS{Publ. Variable Stars Sect. R. Astron. Soc. New Zealand}
\DeclareAbbreviation\PZ{Perem. Zvezdy}
\DeclareAbbreviation\PZP{Perem. Zvezdy Pril.}
\DeclareAbbreviation\QJRAS{QJRAS}
\DeclareAbbreviation\RMxAA{Rev. Mexicana Astron. Astrof.}
\DeclareAbbreviation\RvMA{Reviews of Modern Astron.}
\DeclareAbbreviation\Sci{Science}
\DeclareAbbreviation\SvA{Soviet Astronomy}
\DeclareAbbreviation\SvAL{Soviet Astronomy Letters}
\DeclareAbbreviation\VeSon{Ver\"{o}ff. Sternw. Sonneberg}
\DeclareAbbreviation\VSOLJBul{VSOLJ Variable Star Bull.}
\DeclareAbbreviation\yCat{VizieR Online Data Catalog}
\DeclareAbbreviation\ZA{Z. Astrophys.}

\begin{document}

\SetRunningHead{T. Kato}{Unusual State in V Sagittae}

\Received{}%{yyyy/mm/dd}
\Accepted{}%{yyyy/mm/dd}

\title{Unusual State in V Sagittae}

\author{Taichi \textsc{Kato}}
\affil{Department of Astronomy, Kyoto University,
       Sakyo-ku, Kyoto 606-8502}
\email{tkato@kusastro.kyoto-u.ac.jp}

%%% end:list of authors


\maketitle

\begin{abstract}
   It has been recently demonstrated that the recurring high/low
states in the peculiar binary V Sge can be explained by considering
the limit-cycle oscillation involving negative feedback by the wind on
the mass-transfer from the secondary star.  We noticed, from the
recent observations reported to the VSNET, the presence of a different
state of recurring variations (recurrence time $\sim$ 30 d)
showing a gradual rise and a sudden drop in brightness.
The observed features of
this light variation are unlikely to be reproduced by the limit-cycle
oscillation mechanism involving a mass-transfer variation from
the secondary star (Hachisu, Kato 2003).
We suggest that the phenomenon originates from an unidentified
intrinsic instability in the disk or in the wind of a high luminosity
object as in V Sge.
\end{abstract}

\section{Introduction}

   V Sge is a peculiar variable star which has been receiving outstanding
attention from various researchers.  The short-period binary nature of
V Sge was revealed by \citet{her65vsge}, who classified it as a
``nova-like" variable star mainly from the presence of high excitation
emission lines characteristic to novae and related systems.\footnote{
  Although V Sge is widely recognized as a nova-like variable since
  this discovery of the binary nature, readers should bear in mind that
  the ``nova-like" category of cataclysmic variables (CVs) at that time
  was different (cf. \cite{GCVS3}) from the present one (cf. classification
  scheme from the disk-instability theory: \cite{osa96review});
  the initial ``nova-like" classification did not necessarily imply
  a CV.
}  The nature of the binary components in V Sge was unclear at the
time of \citet{her65vsge}.  More recent observations
(e.g. \cite{wil86vsge}) suggested that the binary contains a compact
accretor, which is most likely a white dwarf.  Although there have
been claims of different interpretations (cf. \cite{loc99vsge}),
the recent detection of the supersoft X-rays (\cite{gre98vsgeSSS};
see also \cite{hoa96puvulfgservsgeROSAT} for a potentially
different state) led to the re-classification of V Sge as
a Galactic luminous supersoft X-ray source (SSXS)
(\cite{gre98vsgeSSS}; \cite{ste98vsgestars};
\cite{pat98vsgetpyx}; \cite{gre00v751cygvsge}).  In particular,
\citet{ste98vsgestars} proposed a new class of variable stars
(V Sge being its prototype) in line with the SSXS interpretation.

   V Sge is known to show long-term light variations characterized
by high and low state, and occasionally with intermediate states
(\cite{sim96vsge}; \cite{rob97vsge}; \cite{sim99vsge}; \cite{sim00vsge}).
Although the origin of this light variation had long been a mystery,
\citet{hac03vsge} recently proposed a mechanism involving time-evolution
of the wind from a luminous white dwarf, time-evolution of the
accretion disk, and negative feedback by the wind on
the mass-transfer from the secondary star.  Although the light curve
of V Sge between the years 1990--1996 seems to have been
phenomenologically well reproduced by this model,
we noticed a significant departure in the
2003 light curve from what is expected from \citet{hac03vsge}.
We thereby report on this observation for future discussions and
theoretical modeling.

\section{Observations}

\begin{figure}
  \begin{center}
%    \FigureFile(88mm,60mm){lc.eps}
    \FigureFile(88mm,60mm){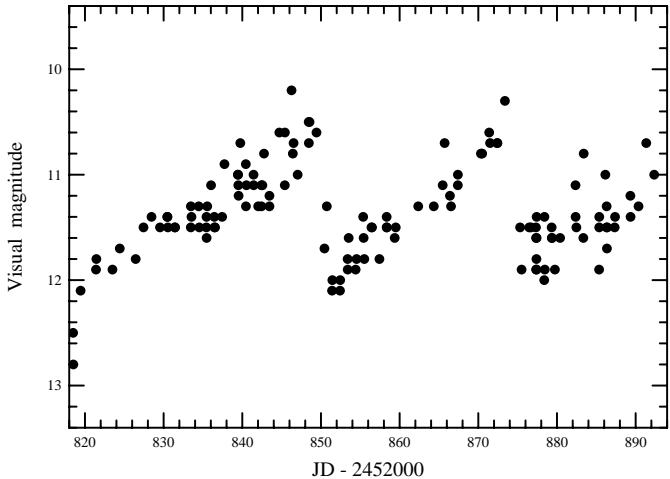}
  \end{center}
  \caption{Light curve of V Sge in 2003 constructed from the visual
  observations reported to the VSNET Collaboration.
  }
  \label{fig:lc}
\end{figure}

   We examined the observations of V Sge posted to VSNET Collaboration
\citep{VSNET}.\footnote{
$\langle$http://www.kusastro.kyoto-u.ac.jp/vsnet/$\rangle$.}
The observations used $V$-band comparison stars, and typical errors
of individual estimates are smaller than 0.3 mag, which would not affect
the following discussion.
Figure \ref{fig:lc} shows a segment of the 2003 light curve of V Sge
showing unusual light variation.  The variation was characterized
by a slow gradual brightening by 1.5 mag, followed by a sudden drop
within 3 d.  This feature seems to recur with a time scale of $\sim$30 d.

\section{Discussion}

   Following the model by \citet{hac03vsge}, the limit cycle oscillation
responsible for the recurrence of low and high states in V Sge involves
the negative feedback of mass-transfer caused by the stripping of the
secondary star by the wind from the primary.  As shown in
\citet{hac03vsge}, this process would naturally require {\it gradual
fading} during the wind phase, in which the mass-transfer from the
secondary is gradually reduced.  The present observation shows the
perfectly contrary sequence of high/low state transitions, showing
a gradual brightening and a sudden drop, with the amplitude comparable
to the high/low state transitions observed previously.
This difference can be best
demonstrated by the comparison with figure 13 in \citet{hac03vsge}.
The recurrence time ($\sim$ 30 d) is also much shorter than what can
be expected from the reasonable parameter ranges presented in
\citet{hac03vsge}.  This observation suggests that at least some stage
of variations in V Sge is not driven by the limit-cycle oscillation
involving the mass-transfer variation from the secondary star.

   V Sge displayed an evolution of its activity over
the last $\sim$70 years \citep{sim99vsge} and switched from one type of
photometric activity to another on the time scale of years several times.
The  behavior in figure \ref{fig:lc} then represents appearance of
another type.

   Considering the present noticeable departure of V Sge
itself from the model light curves and the short transition time scale,
a different mechanism should be sought, including the intrinsic
instability in the disk or wind in a high-luminosity object.

   It would be worth noting that the symbiotic star CH Cyg occasionally
shows high/low state transitions similar to that of V Sge
\citep{kat00chcyg}.  If the variation in CH Cyg is indeed analogous to
V Sge-type high/low state transitions, a mechanism involving
the mass-transfer variation from the secondary star would not be
able to explain the observed time scale in such a long-period binary
as in CH Cyg.  An alternative (although not yet identified) mechanism
in V Sge might be a clue to understanding the common underlying
physics among a wide diversity of high-luminosity accreting systems
ranging from V Sge stars to symbiotic binaries.  It would be worth
noting that a wind-driven mechanism has been proposed for
variable mass-accretion rate in symbiotic binaries \citep{bis02zand},
which may be also applicable to V Sge with strong winds
(e.g. \cite{loc99vsge}).

\vskip 3mm

   We are grateful to many amateur observers for supplying their vital
visual and CCD estimates via VSNET.  The author is grateful to an anonymous
referee for drawing attention to the work by Bisikalo et al.
This work is partly supported by a grant-in-aid (13640239, 15037205)
from the Japanese Ministry of Education, Culture, Sports, Science and
Technology.

\end{document}